\documentclass[conference]{IEEEtran}
\usepackage[T1]{fontenc}
\usepackage{graphicx}
\usepackage[nobreak]{cite}
\usepackage{url}
\usepackage{xspace}

\newcommand{\Heading}[1]{\textbf{#1.}}
\newcommand{\Refactoring}[1]{\textsf{#1}}
\newcommand{\Param}[1]{\textit{#1}}
\newcommand{\Type}[1]{\textit{#1}}

\begin{document}

\title{RefactorHub: A Commit Annotator for Refactoring}

\author{\IEEEauthorblockN{Ryo Kuramoto, Motoshi Saeki, and Shinpei Hayashi}
\IEEEauthorblockA{%
  \textit{School of Computing, Tokyo Institute of Technology}\\
  Meguro-ku, Tokyo 152--8550, Japan\\
  kuramoto@se.c.titech.ac.jp, \{saeki,hayashi\}@c.titech.ac.jp}
}

\maketitle

\begin{abstract}
It is necessary to gather real refactoring instances while conducting empirical studies on refactoring.
However, existing refactoring detection approaches are insufficient in terms of their accuracy and coverage.
Reducing the manual effort of curating refactoring data is challenging in terms of obtaining various refactoring data accurately.
This paper proposes a tool named RefactorHub, which supports users to manually annotate potential refactoring-related commits obtained from existing refactoring detection approaches to make their refactoring information more accurate and complete with rich details.
In the proposed approach, the parameters of each refactoring operation are defined as a meaningful set of code elements in the versions before or after refactoring.
RefactorHub provides interfaces and supporting features to annotate each parameter, such as the automated filling of dependent parameters, thereby avoiding wrong or uncertain selections.
A preliminary user study showed that RefactorHub reduced annotation effort and improved the degree of agreement among users.
Source code and demo video are available at \url{https://github.com/salab/RefactorHub}
\end{abstract}


\section{Introduction}\label{c:introduction}

Refactoring is a technique to improve the internal structure of software while preserving their external behavior \cite{Fowler_RefactoringImprovingDesign_1999}.
Many refactoring tools have been developed, and refactoring is now common in modern software development.
Also, many empirical studies about refactoring have been conducted using practical refactoring instances, such as investigations on the relationship between refactorings and the merging conflicts \cite{Mahmoudi_AreRefactoringsBlame_2019}, the characteristics of \Refactoring{Extract Method} \cite{Hora_CharacteristicsMethodExtractions_2020}, or the reason developers refactor code \cite{Silva_WhyWeRefactor_2016}.

When conducting empirical studies, it is necessary to collect data on refactoring instances that were applied in practice.
There are four pre-requisites in terms of collecting data on refactoring instances: sufficient number, accuracy, coverage in terms of their types, and detailed information needed for the studies.
If the collected data is insufficient, it may not be possible to draw a solid conclusion.
If the accuracy of the collected data is low, i.e., if it includes many false positives, analyses conducted using such data may lead to a faulty conclusion.
Also, depending on the type of study, one may need refactoring information on particular parameters of particular refactorings.

There are several ways to collect refactoring instances, such as the use of refactoring detection tools or existing datasets.
For example, refactoring detection tools such as RefDiff~\cite{Silva_RefDiffDetectingRefactorings_2017, Silva_RefDiffMultilanguageRefactoring_2020} or RefactoringMiner~\cite{Tsantalis_AccurateEfficientRefactoring_2018,Tsantalis_RefactoringMiner_2020} can be used to collect a large number of refactoring instances because they can automatically detect refactorings from history.
However, because their accuracy is not 100\%, there is a possibility that some false positive refactorings may be incorrectly detected.
Also, they can only detect certain refactoring types.
When using existing datasets, such as the one proposed by Silva et al.~\cite{Silva_WhyWeRefactor_2016}, users expect them to be highly accurate and sufficiently large, as they have been manually validated and used in previous studies.
However, the refactoring types included in the datasets and the level of detail of each instance vary depending on the nature of the study, and they may be insufficient to be used in another study.
Manual dataset preparation can fulfill the requirements of accuracy, type coverage, and the level of detail; however, a significant amount of effort is necessary to collect the large number of instances required.

In this paper, we propose a tool named RefactorHub, which supports users to manually \emph{annotate} potential refactoring-related commits obtained from existing refactoring detection approaches that may be inaccurate and/or less informative to make their refactoring information more accurate and complete with rich details.
In RefactorHub, the parameters of each refactoring operation, which define where the refactoring was performed and which code elements were produced and/or changed, are defined as a meaningful set of code elements in the versions before and after refactoring. 
RefactorHub provides interfaces and supporting features to annotate each parameter by selecting and completing code elements on a commit, thereby avoiding wrong or uncertain selections.


\section{Annotation-Based Approach\\for Collecting Refactoring Data}\label{c:approach}

\begin{figure}[tb]\centering
  \includegraphics[width=\linewidth]{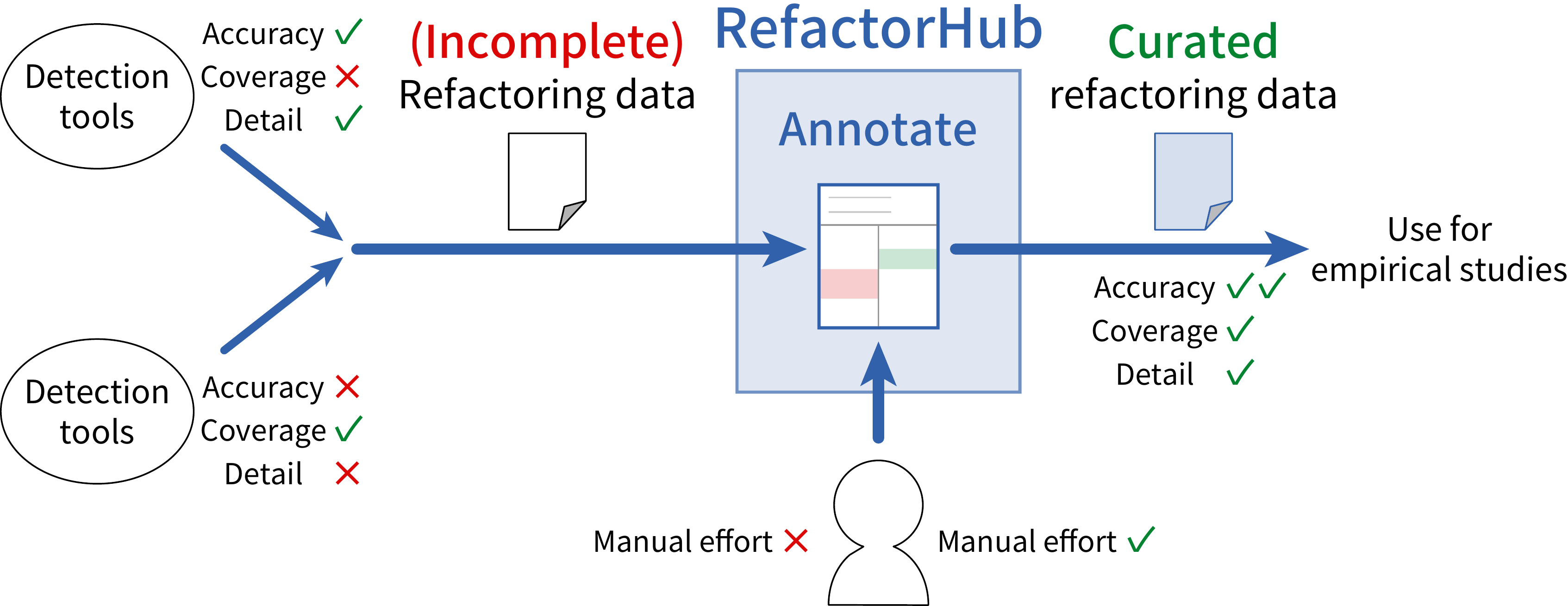}
  \caption{Annotation-based approach to collect refactoring data.}\label{f:overview}
\end{figure}

To solve the issue of collecting refactoring data, we follow the approach shown in Fig.~\ref{f:overview}.
In our approach, we first take incomplete refactoring data as input.
They can be obtained from existing refactoring detection tools/approaches or from existing datasets.
Subsequently, refactoring instances are manually verified and/or completed using a change annotation environment.
We call this manual verification and completion process \emph{annotation}.
RefactorHub is a tool that supports this annotation process.
The final output of the annotation process is the refined refactoring instance data, which are intended to be used in empirical studies on refactoring.
  
The approach shown in Fig.~\ref{f:overview} aims to solve issues such as accuracy, coverage, degree of detail, and the insufficient amount of refactoring data obtained from existing detection tools and datasets.
As mentioned above, when using refactoring detection tools or existing datasets, we can expect to obtain a sufficient number of candidate refactoring instances.
However, such approaches may have accuracy, type coverage, or level of detail issues.
Therefore, we compensate each weakness by manually supplementing accuracy and rich details.
Because this manual validation and completion process is time-consuming, we attempt to mitigate it with RefactorHub, a tool that provides a change annotation environment.

The accuracy and coverage of refactoring detection vary depending on individual approaches.
On the one hand, several tools such as RefactoringMiner \cite{Tsantalis_AccurateEfficientRefactoring_2018,Tsantalis_RefactoringMiner_2020} or RefDiff \cite{Silva_RefDiffDetectingRefactorings_2017,Silva_RefDiffMultilanguageRefactoring_2020} can detect refactorings with very high accuracy.
In addition, the latest version of RefactoringMiner \cite{Tsantalis_RefactoringMiner_2020} supports JSON output with rich details of refactoring operations.
However, both tools can only detect certain refactoring types.
On the other hand, more lightweight approaches such as keyword search of commit messages \cite{ratzinger-msr-2008,stroggylos-wosq-2007,alomar-iwor2019} may be able to find refactoring-related commits that cannot be detected by the accurate tools mentioned above; however, they are less accurate and do not provide detailed location information of the refactorings.
RefactorHub can be used for two purposes.
The first is to verify the detection results of the aforementioned accurate tools.
The second is to identify the refactoring type and parameters from the changes in the candidate commits obtained from the latter lightweight approaches.

\begin{figure}[tb]\centering
  \includegraphics[width=\linewidth]{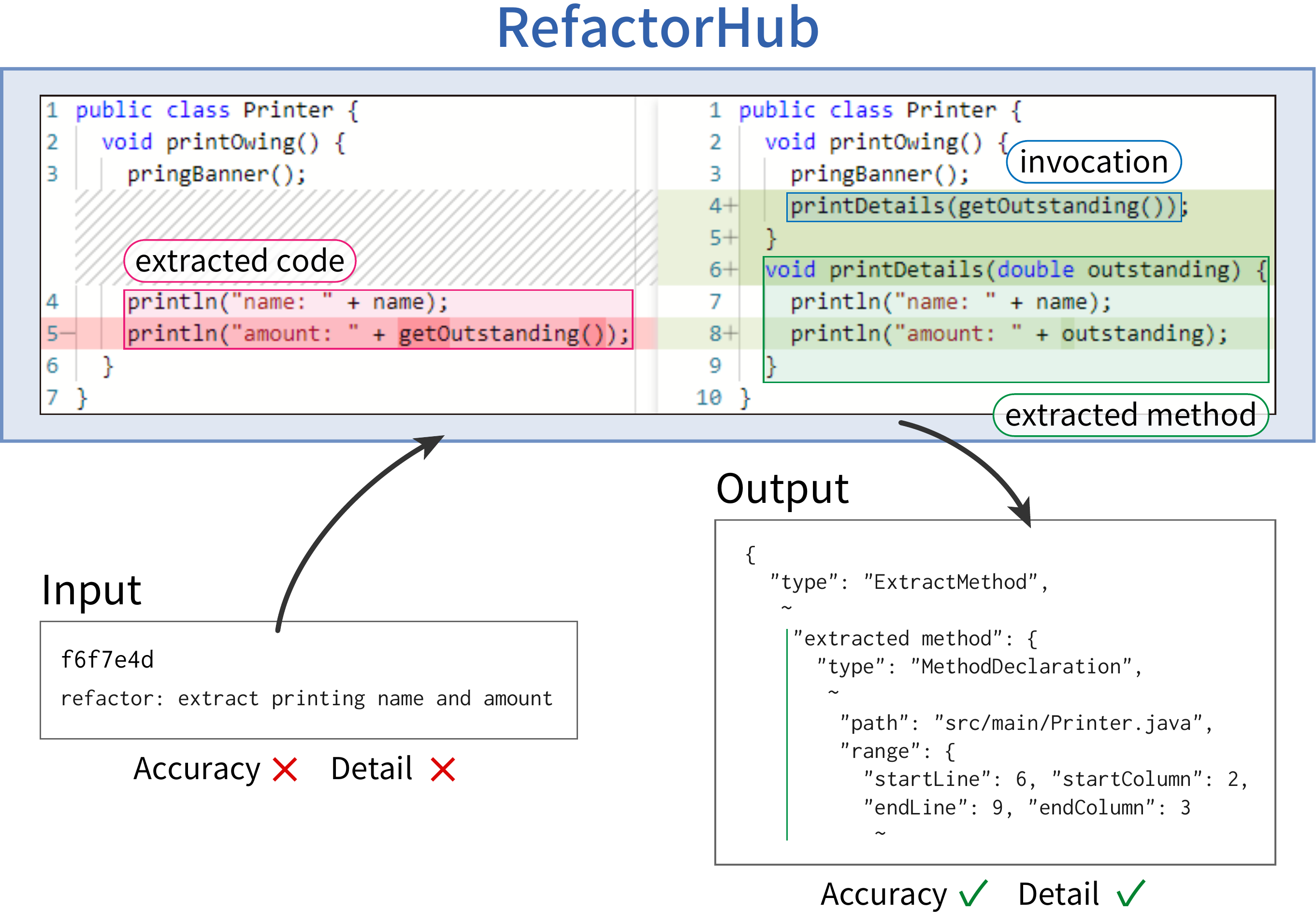}
  \caption{Annotating \Refactoring{Extract Method}.}\label{f:annotation}
\end{figure}

Figure~\ref{f:annotation} shows the outline of annotating the \Refactoring{Extract Method} using RefactorHub\@.
In the context of RefactorHub, the annotation process selects code elements related to the target refactoring on the source code before and after changes in a commit.
First, we use as inputs a commit that may have been refactored and the hint for the refactoring, which can be obtained from refactoring detection tools or existing datasets.
The hint can be a text description about the refactoring obtained from detection tools or from the commit message.
It can also be an initial annotation result if the used detection tools can output such information.
In Fig.~\ref{f:annotation}, a commit with a refactoring description from Refactoring Oracle~\cite{refactoring-oracle} is used as the input.
Next, the user manually selects code elements related to the refactoring from the source code before and after the commit and attaches their meaning to the elements by naming them as used in further empirical studies.
In Fig.~\ref{f:annotation}, for the \Refactoring{Extract Method} applied in the commit, the extracted code before refactoring is selected as \Param{extracted code}, the newly defined extracted method after refactoring is selected as \Param{extracted method}, and the part replaced by a method invocation is selected as \Param{invocation}.
Finally, the set of parameters with the refactoring name is output as a set of meaningful code elements in the source code before and after the refactoring.

\section{RefactorHub in a Nutshell}\label{c:implementation}

\begin{figure*}[tb]\centering
  \includegraphics[width=\linewidth]{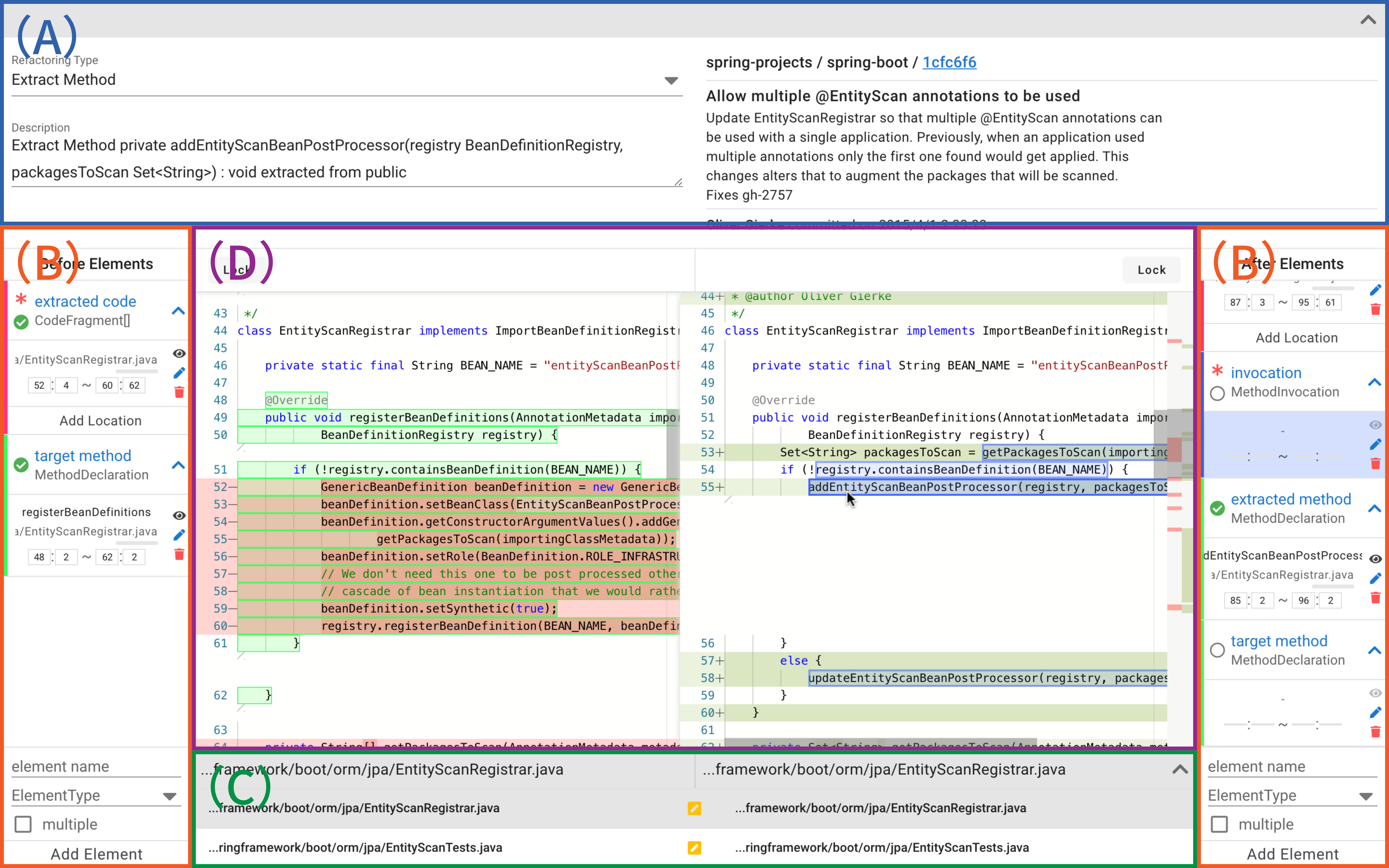}
  \caption{Screenshot of RefactorHub.}\label{f:screen}
\end{figure*}

\subsection{Overview}
RefactorHub is a commit annotation environment for validating and completing given refactoring data to obtain detailed and accurate data.
RefactorHub is implemented as a web application, and it takes a JSON file representing the input refactoring data and outputs another JSON file containing the curated refactoring data.

Figure~\ref{f:screen} show a screenshot of RefactorHub.
In this figure, the user opens a commit that may include \Refactoring{Extract Method} and describing its details.
Pane~(A) shows the metadata of the given refactoring and the commit to which it belongs.
Pane~(B) shows the list of the parameters of the given refactoring; the parameters related to the code fragments before and after refactoring are listed on the left and right sides, respectively.
Pane~(C) lists all the files that have been changed in the target commit.
Pane~(D) shows the source code differences of the files selected in Pane~(C).
Users can select the code elements related to refactoring in the differences.
In a typical annotation flow, the user first reads the information from Pane (A), understands the refactoring applied in the commit by selecting the files in (C) and showing them in (D), and determines the code elements in (D) as the parameters listed in (B).

\subsection{Refactoring Type}
To increase the generality of RefactorHub, each parameter of refactoring operations in the output is defined as a set of text ranges of source code.
However, to mitigate errors due to inconsistency among users, such as wrong selection and/or blur in the selected range in the annotated results, we use syntactic code elements in an abstract syntax tree (AST) for supportively representing refactoring parameters.
Parameters are typed using the categories of AST nodes, each of which represents a code element in the source code.
For example, the type \Type{FieldDeclaration} represents a field declaration, and \Type{MethodInvocation} represents a method invocation.
A refactoring type is defined as a set of parameters, and each of them represents a syntactic code element in the source code before or after the changes in the commit.
For example, in defining \Refactoring{Extract Method}, its parameter \Param{extracted method} represents the newly defined method in the source code after refactoring, and this parameter is typed as \Type{MethodDeclaration}.
Note that a special type \Type{CodeFragment} can also be used to specify any text range in a method regardless of its type, e.g., the extracted code range in \Refactoring{Extract Method}.
Currently, 25 refactoring types are predefined, and users can additionally define their own refactoring types.

The annotation process using the defined refactoring types can reduce the bias of the annotated results.
The use of refactoring types can prevent situations where the annotation results obtained by two users for the same parameter do not match with each other because of their misunderstanding of the expected type of code element to be selected.
Moreover, AST-based typing enables users to avoid ambiguities in the selected ranges when specifying a code element in source code.
For example, when specifying the text range of a method declaration, the inclusion of its surrounding white spaces and comments, which are non-essential for the method, is ambiguous if the range can be freely selected.
Validation of the selected text range for a parameter according to the parameter type can prevent annotators from selecting an unexpectedly wrong or ununified selection.

\subsection{Features}

\subsubsection{Selection Support}
RefactorHub can automatically fill refactoring parameters by selecting code elements in the differences in a commit, thereby reducing manual effort.
The tool parses the source files shown in Pane~(D) and creates their ASTs.
When a user starts to annotate for a specific parameter, the tool specifies the parameter's type and enumerates matched nodes of the specified type in the ASTs.
The matched nodes are highlighted in the differences shown in Pane (D).
Upon clicking one of the highlighted nodes, the tool automatically fills the contents of the parameter.

\subsubsection{Autofill}
The \emph{autofill} feature in RefactorHub automatically derives the value of some refactoring parameters from another parameter, which reduces users' manual effort.
For example, \Refactoring{Move Field} has two parameters, \Param{moved field} (the moved field declaration) and \Param{references}~(identifiers that refer to the moved field).
Once a user determines the \Param{moved field} parameter, the contents of the \Param{references} parameter are automatically derived as a list of identifiers that occur on the code difference, which may refer to the specified field declaration.
This derivation is based on the equality of the identifier names.
In addition, an ancestor parameter is derivable from a given parameter, e.g., \Param{extracted method} parameter (the method declaration containing the code fragment to be extracted) in \Refactoring{Extract Method} can be derived as a parent method declaration of the \Param{extracted code} parameter (the code fragment removed by extraction).
This autofill feature does not aim at detailed derivation for particular refactorings but a \emph{generic} derivation for any refactoring so that it is not aware of the semantics of the refactorings, which may lead to an incorrect derivation.
Users are expected to validate the derivation results manually and correct them if necessary.


\section{Preliminary User Study}\label{c:evaluation}

A preliminary user study was conducted with four participant developers (their Java experience: 1.5--4 years) to evaluate the usefulness of RefactorHub in terms of its saved efforts and improved agreement.
Regarding the inputs, we extracted four commits from Refactoring Oracle~\cite{refactoring-oracle}, which is a well-known refactoring dataset, for each of the four refactoring types: \Refactoring{Extract Method}, \Refactoring{Move Field}, \Refactoring{Move Class}, and \Refactoring{Rename Variable}, resulting in a total of 16 refactoring instances.
We asked four participants to annotate these commits with the given information on refactoring types.
To confirm the effectiveness of RefactorHub's \emph{autofill} feature for each refactoring instance, two of the four participants were assigned to annotate it with the feature available, whereas the other two annotated the version without the feature.
Note that we tried to avoid imbalanced assignments in terms of the refactoring types and availability of the autofill feature.

\begin{figure}[tb]\centering
    \includegraphics[width=\linewidth]{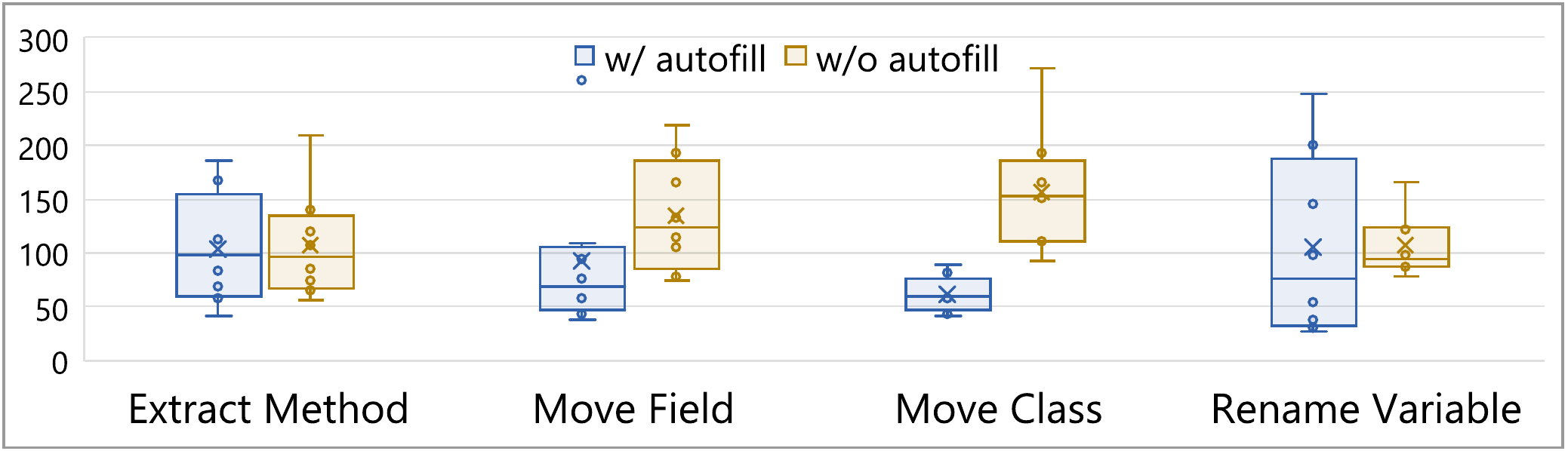}
    \caption{Time spent in the annotation per refactoring type (sec).}\label{f:times}
\end{figure}
\Heading{Annotation time spent}
Figure~\ref{f:times} compares the annotation time for each refactoring type with and without the autofill feature available.
The annotation time is defined as the difference between the time the subject starts annotating any parameter and the time the subject finishes annotating the last element.
As shown in the figure, in \Refactoring{Move Field} and \Refactoring{Move Class}, the annotation time was shorter when the autofill feature was available.
More specifically, the average annotation time for the \Param{references} parameter before applying \Refactoring{Move Class}, with and without the automatic derivation feature, were 10.2 and 68.0 secs, respectively.
This might be because the advantage of the autofill feature was greater for elements that select a wide range of references.
In contrast, the difference in \Refactoring{Rename Variable} was small, although it also involves the parameter \Param{references}.
This might be because the range of the \Param{references} in \Refactoring{Rename Variable} is limited within the method scope, which causes the effect of the autofill feature to be limited.

\Heading{Agreement level}
We computed the inter-user agreement rate, which tells how many refactoring parameters were exactly matched.
If the range of a parameter annotated by a user was exactly matched to that annotated by another user, we considered them as matched.
The average inter-user agreement rate was 87.5\% when the autofill feature was available, and it was higher than 62.5\% when the feature was unavailable.

\begin{figure}[tb]\centering
  \includegraphics[width=\linewidth]{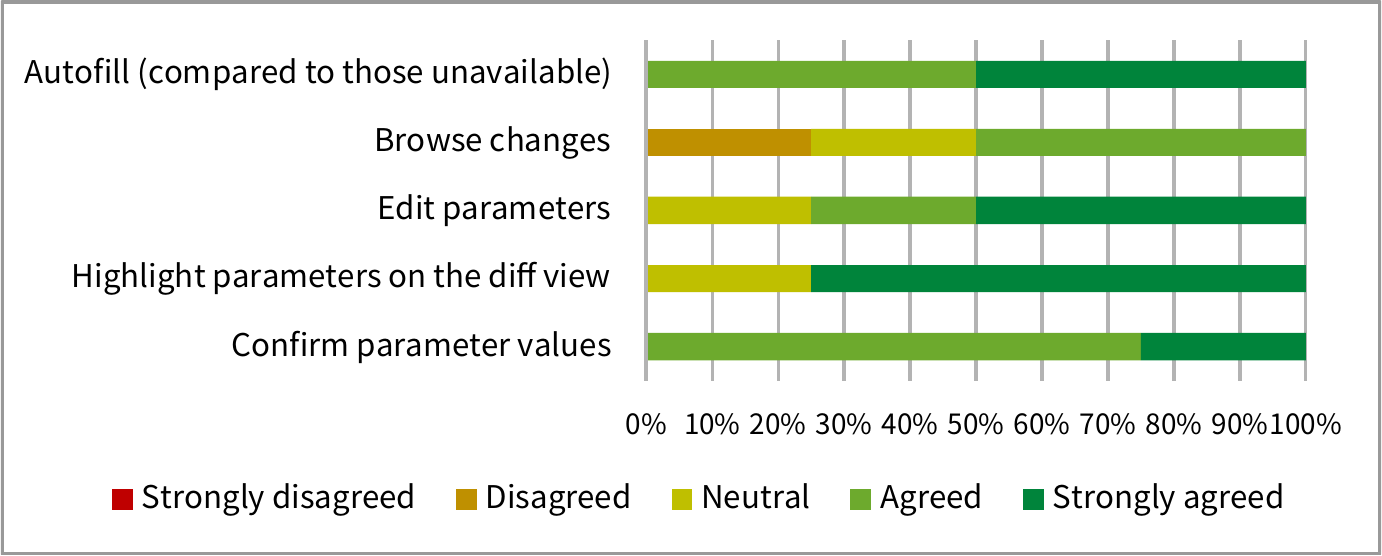}
  \caption{How much were the feature of RefactorHub useful?}\label{f:enquete_1}
\end{figure}
\Heading{Usability}
We asked the four participants whether they found RefactorHub features useful. The results are shown in Fig.~\ref{f:enquete_1}.
The participants responded positively to all the features.
In particular, two of the four participants strongly agreed, and the other agreed on the usefulness of the autofill feature.

Although this user study involved only four participants, four refactoring types, and 16 refactoring instances,
these results suggest that RefactorHub could reduce the time spent to annotate with a higher agreement.


\section{Conclusion}\label{c:conclusion}

In this paper, we propose RefactorHub, which verifies and complements input refactoring data obtained from refactoring detection tools and/or existing datasets to make them more detailed and accurate.
In RefactorHub, refactoring operations are defined as a set of syntactic elements in AST as their parameters, and the parameters are annotated on the source code difference of the given commit according to their types.
A preliminary user study has shown that the supportive features of RefactorHub was effective in annotating refactoring data.

Future work can be listed up as follows.
\begin{itemize}
  \item Improving the \emph{autofill} feature.
        Making it more accurate and increasing the availability coverage of this feature enables users to annotate refactoring data more effectively.
  \item Supporting compound refactorings.
        Studies showed that refactorings are frequently applied in meaningful combinations within a single commit~\cite{Sousa_CharacterizingIdentifyingComposite_2020}.
        It would be useful to annotate multiple refactorings in a single commit.
  \item Realizing an open platform for sharing refactoring datasets.
        RefactorHub, as its name implies, aims to be an open dataset platform by linking refactoring data.
        Similar to Landfill~\cite{Palomba_LandfillOpenDataset_2015} for code smells and InspectorClone~\cite{Saini_AutomatingPrecisionStudies_2019} for code clones, we would like to build a platform that enables users to share datasets of refactoring instances and allows users to refine them.
\end{itemize}

\section*{Acknowledgments}
This work was partly supported by JSPS KAKENHI JP18K11238.

\IEEEtriggeratref{9}

\begin{thebibliography}{10}
\providecommand{\url}[1]{#1}
\csname url@samestyle\endcsname
\providecommand{\newblock}{\relax}
\providecommand{\bibinfo}[2]{#2}
\providecommand{\BIBentrySTDinterwordspacing}{\spaceskip=0pt\relax}
\providecommand{\BIBentryALTinterwordstretchfactor}{4}
\providecommand{\BIBentryALTinterwordspacing}{\spaceskip=\fontdimen2\font plus
\BIBentryALTinterwordstretchfactor\fontdimen3\font minus
  \fontdimen4\font\relax}
\providecommand{\BIBforeignlanguage}[2]{{%
\expandafter\ifx\csname l@#1\endcsname\relax
\typeout{** WARNING: IEEEtran.bst: No hyphenation pattern has been}%
\typeout{** loaded for the language `#1'. Using the pattern for}%
\typeout{** the default language instead.}%
\else
\language=\csname l@#1\endcsname
\fi
#2}}
\providecommand{\BIBdecl}{\relax}
\BIBdecl

\bibitem{Fowler_RefactoringImprovingDesign_1999}
M.~Fowler, \emph{Refactoring: Improving the Design of Existing Code}, ser.
  Addison {{Wesley}} Object Technology Series.\hskip 1em plus 0.5em minus
  0.4em\relax {Addison-Wesley}, 1999.

\bibitem{Mahmoudi_AreRefactoringsBlame_2019}
M.~Mahmoudi, S.~Nadi, and N.~Tsantalis, ``Are refactorings to blame? {An}
  empirical study of refactorings in merge conflicts,'' in \emph{Proceedings of
  the 26th IEEE International Conference on Software Analysis, Evolution and
  Reengineering (SANER 2019)}, 2019, pp. 151--162.

\bibitem{Hora_CharacteristicsMethodExtractions_2020}
A.~Hora and R.~Robbes, ``\BIBforeignlanguage{en}{Characteristics of method
  extractions in {Java}: A large scale empirical study},''
  \emph{\BIBforeignlanguage{en}{Empirical Software Engineering}}, vol.~25,
  no.~3, pp. 1798--1833, 2020.

\bibitem{Silva_WhyWeRefactor_2016}
D.~Silva, N.~Tsantalis, and M.~T. Valente, ``Why we refactor? {C}onfessions of
  {GitHub} contributors,'' in \emph{Proceedings of the 24th ACM SIGSOFT
  International Symposium on Foundations of Software Engineering (FSE 2016)},
  2016, pp. 858--870.

\bibitem{Silva_RefDiffDetectingRefactorings_2017}
D.~Silva and M.~T. Valente, ``{RefDiff}: Detecting refactorings in version
  histories,'' in \emph{Proceedings of the 14th International Conference on
  Mining Software Repositories (MSR 2017)}, {Buenos Aires, Argentina}, 2017,
  pp. 269--279.

\bibitem{Silva_RefDiffMultilanguageRefactoring_2020}
D.~Silva, J.~Silva, G.~J. De~Souza~Santos, R.~Terra, and M.~T.~O. Valente,
  ``{RefDiff} 2.0: {A} multi-language refactoring detection tool,'' \emph{IEEE
  Transactions on Software Engineering}, 2020, doi: 10.1109/TSE.2020.2968072.

\bibitem{Tsantalis_AccurateEfficientRefactoring_2018}
N.~Tsantalis, M.~Mansouri, L.~Eshkevari, D.~Mazinanian, and D.~Dig, ``Accurate
  and efficient refactoring detection in commit history,'' in \emph{Proceedings
  of the 40th IEEE/ACM International Conference on Software Engineering (ICSE
  2018)}, 2018, pp. 483--494.

\bibitem{Tsantalis_RefactoringMiner_2020}
N.~Tsantalis, A.~Ketkar, and D.~Dig, ``{RefactoringMiner} 2.0,'' \emph{IEEE
  Transactions on Software Engineering}, 2020, doi: 10.1109/TSE.2020.3007722.

\bibitem{ratzinger-msr-2008}
J.~Ratzinger, T.~Sigmund, and H.~C. Gall, ``On the relation of refactorings and
  software defect prediction,'' in \emph{Proceedings of the 5th Working
  Conference on Mining Software Repositories (MSR 2008)}, 2008, pp. 35--38.

\bibitem{stroggylos-wosq-2007}
K.~Stroggylos and D.~Spinellis, ``{Refactoring}--does it improve software
  quality?'' in \emph{Proceedings of the 5th International Workshop on Software
  Quality (WoSQ 2007)}, 2007.

\bibitem{alomar-iwor2019}
E.~{AlOmar}, M.~W. {Mkaouer}, and A.~{Ouni}, ``Can refactoring be
  self-affirmed? {An} exploratory study on how developers document their
  refactoring activities in commit messages,'' in \emph{Proceedings of the 3rd
  IEEE/ACM International Workshop on Refactoring (IWoR 2019)}, 2019, pp.
  51--58.

\bibitem{refactoring-oracle}
N.~Tsantalis, M.~Mansouri, L.~M. Eshkevari, and D.~Mazinanian, ``Refactoring
  oracle,'' \url{http://refactoring.encs.concordia.ca/oracle/}, 2020.

\bibitem{Sousa_CharacterizingIdentifyingComposite_2020}
L.~Sousa, D.~Cedrim, A.~Garcia, W.~Oizumi, A.~C. Bibiano, D.~Oliveira, M.~Kim,
  and A.~Oliveira, ``Characterizing and identifying composite refactorings:
  {C}oncepts, heuristics and patterns,'' in \emph{Proceedings of the 17th
  International Conference on Mining Software Repositories (MSR 2020)}, 2020,
  pp. 186--197.

\bibitem{Palomba_LandfillOpenDataset_2015}
F.~Palomba, D.~Di~Nucci, M.~Tufano, G.~Bavota, R.~Oliveto, D.~Poshyvanyk, and
  A.~De~Lucia, ``{Landfill}: {An} open dataset of code smells with public
  evaluation,'' in \emph{Proceedings of the 12th IEEE/ACM Working Conference on
  Mining Software Repositories (MSR 2015)}, 2015, pp. 482--485.

\bibitem{Saini_AutomatingPrecisionStudies_2019}
V.~Saini, F.~Farmahinifarahani, Y.~Lu, D.~Yang, P.~Martins, H.~Sajnani,
  P.~Baldi, and C.~V. Lopes, ``Towards automating precision studies of clone
  detectors,'' in \emph{Proceedings of the 41st IEEE/ACM International
  Conference on Software Engineering (ICSE 2019)}, 2019, pp. 49--59.

\end{thebibliography}


\end{document}